\documentclass[10pt,A4paper,conference]{IEEEtran}

\begin{document}
\title{Universal Codes as a Basis for  Nonparametric Testing of Serial Independence for Time Series
}
\author{\authorblockN{Boris Ryabko\authorrefmark{1} and Jaakko Astola\authorrefmark{2}}
\authorblockA{\authorrefmark{1}Institute of Computational Technology of
Siberian Branch of Russian Academy of Science,
\\
}
\authorblockA{\authorrefmark{2}Tampere University of Technology, Finland,
\\
Email: boris@ryabko.net, jaakko.astola@tut.fi} } \maketitle

\begin{abstract}
We  consider a stationary and ergodic source $p$ generated symbols
$x_1 \ldots x_t$ from some finite set $A$ and a null hypothesis
$H_0$ that $p$ is Markovian source with memory (or connectivity)
not larger than $m,\: (m \geq 0).$ The alternative hypothesis
$H_1$ is that the sequence is generated by a stationary and
ergodic source, which differs from the source under $H_0$. In
particular, if $m= 0$ we have the null hypothesis $H_0$ that the
sequence is generated by Bernoully source (or the hypothesis that
$x_1 \ldots x_t$ are independent.) Some new tests which are based
on universal codes and universal predictors, are suggested.
\end{abstract}


\section{Introduction}

Nonparametric testing for independence of time series is very
important in statistical applications. There is an extensive
literature dealing with nonparametric independence testing; a
quite full review can be found in \cite{GKR}.

In this paper,  we consider a source (or process), which generates
elements from a finite set $A$ and two following  hypotheses:
$H_0$ is that the source is Markovian one, which memory (or
connectivity) not larger than $m,\: (m \geq 0),$ and the
alternative hypothesis $H_1$ that the sequence is generated by a
stationary and ergodic source, which differs from the source under
$H_0$. The testing should be based on a sample $x_1 \ldots x_t$
generated by the source.

For example, the sequence $x_1 \ldots x_t$ might be a DNA-string
and one can consider the question about the depth of the
statistical dependence.

We suggest a family of tests that are based on so called universal
predictors (or universal data compression methods). The Type I
error of the suggested tests is not larger than a given $\alpha
\:( \alpha \in (0,1)\, ) $ for any source under $H_0,$ whereas the
Type II error  for any source under $H_1$  tends to 0, when the
sample size $t$ grows.

The suggested tests are based on results and ideas of Information
Theory and, especially, those of the universal coding theory.
Informally, the main idea of the tests can be described as
follows. Suppose that the source generates letters from an
alphabet $A$ and one wants to test $H_0$ (the source is Morkovian
with memory $m, m \geq 0.$ ) First we recall that there exist
universal codes which, informally speaking can "compress" any
sequence generated by a stationary and ergodic source, to the
length $t h_\infty$ bits, where $ h_\infty$ is the limit Shannon
entropy and $t$ tends to infinity. Second, it is well known in
information theory that  $h_\infty $ equals $m$th-order
(conditional) Shannon entropy $h_{m}$, if $H_0$ is true,  and
$h_\infty$ is strictly less than $h_{m}$ if $H_1$ is true. So, the
following test looks like natural: Compress the sample sequence
$x_1 \ldots x_t$ by a universal code and compare the lengths of
the obtained file with  $t h^*_m, $ where $h^*_m$ is an estimate
of $h_m.$ If the length of the compressed file is significantly
less than $t h^*_m, $ then the hypothesis $H_0$ should be
rejected.

This is no surprise that the results and ideas of a universal
coding theory can be applied to some classical problems of
mathematical statistics. In fact, methods of universal coding (and
a closely connected  universal prediction) are intended to extract
information from observed data in order to compress (or predict)
data efficiently in a case where the source statistics  is
unknown. Recently such a connection between universal coding and
mathematical statistics was  used in \cite{CS} for estimating the
order of Markov sources and  for constructing efficient tests for
randomness, i.e. for testing the hypothesis $\hat{H}_0$ that a
sequence is generated by a Bernoulli source and all letters have
equal probabilities against $\hat{H}_1$ that the sequence is
generated by a stationary and ergodic source, which differs from
the source under $\hat{H}_0,$ see \cite{R-M}.

The outline of the paper is as follows. The next part  contains
definitions and necessary  information  from the theory of
universal coding and universal prediction. Part three is devoted
to the testing of  the above described hypotheses. All proofs are
given in the appendix.

\section{Definitions and Preliminaries.}

Consider an alphabet $A=\{a_1,\cdots,a_n\}$ with $n\geq2$ letters
and denote by $A^t$ the set of words $x_1\cdots x_t$ of length $t$
from $A$. Let $\mu$ be a source which generates letters from $A$.
Formally, $\mu$ is a probability distribution on the set of words
of infinite length or, more simply, $\mu=(\mu^t)_{t\geq1}$ is a
consistent set of probabilities over the sets $A^t\,;\;t\geq1$. By
$M_\infty(A)$ we denote the set of all stationary and ergodic
sources, which generate letters from $A.$ Let $M_k(A) \subset
M_\infty(A)$ be the set of Markov sources with memory (or
connectivity) $k, \, k\geq 0. $ More precisely, by definition $\mu
\in M_k(A)$ if
$$
\mu (x_{t+1} = a_{i_1} / x_{t} = a_{i_2}, x_{t-1} = a_{i_3},\,
...\,, x_{t-k+1} = a_{i_{k+1}},\, ... ) $$
$$  = \mu (x_{t+1} = a_{i_1} / x_{t}  = a_{i_2}, x_{t-1} = a_{i_3},\, ...\,, x_{t-k+1} = a_{i_{k+1}})
$$ for all $t \geq k $ and $a_{i_1}, a_{i_2}, \ldots \, \in A.$ By
definition, $M_0(A)$ is the set of all Bernoulli (or i.i.d.)
sources over $A$.

\newpage

\textbf{2.1 Universal prediction.}

Now we briefly describe results and methods of universal coding
and prediction, which will be used later. Let a source generate a
message $x_1\ldots x_{t-1}x_t\ldots $ and let $ \nu^t(a)$ denote
the count of letter $a$ occurring in the word $x_1\ldots
x_{t-1}x_t $. After the
 first $t$ letters $x_1,\ldots, x_{t-1},x_t$ have been processed
the following letter $ x_{t+1}$ needs to be predicted. By
definition, a prediction is a set of non-negative numbers
$\pi(a_1|x_1\cdots x_t),\cdots, \pi(a_n|x_1\cdots x_t)$ which are
estimates of the unknown conditional probabilities
$p(a_1|x_1\cdots x_t), \cdots,p(a_n|x_1\cdots x_t)$, i.e. of the
probabilities $p(x_{t+1}=a_i| x_1\cdots x_t)$; $i=1,\cdots,n$.

Laplace suggested the following predictor:
\begin{equation}\label{L}
L(a|x_1\cdots x_t) = (\nu^t(a) +1 )/ (t+ |A | ),
\end{equation}
 where
 $|A |$ is the number of letters in the alphabet $A,$ see \cite{FE}.
For example, if $ A= \{0, 1 \}, \: x_1 ... x_5 = 01010,$ then the
Laplace prediction is as follows: $L(x_{6}=0| 01010) = (3+1)/
(5+2) = 4/7, L_0(x_{6}=1 | 01010) = (2+1)/ (5+2) = 3/7.$

In Information Theory  the error of prediction often is estimated
by the the Kullback-Leibler (K-L) divergence between a
distribution $p$ and its estimation. Consider a source $p$ and a
predictor $\gamma$. The {\it{ error }} is characterized by the
divergence
\begin{equation}\label{r0}
\rho_{\gamma,p}(x_1\cdots x_t)= \sum_{a\in A}p(a|x_1\cdots
x_t)\log\frac{p(a|x_1\cdots x_t)} {\gamma(a|x_1\cdots x_t)}.
\end{equation}
(Here and below $\log \equiv \log_2$.) It is well known that for
any distributions  $p$ and $\gamma$ the K-L divergence is
nonnegative and equals 0 if and only if $p(a) = \gamma(a)$ for all
$a,$ see, for ex., \cite{Ga}, that is why the K-L divergence is a
natural estimate of the prediction error. For fixed $t$,
$\rho_{\gamma,p}$ is a random variable, because $x_1, x_2, \cdots,
x_t$ are random variables. We define the average error
  at time $t$ by
$$
\rho^t(p\|\gamma)=E\,\left(\rho_{\gamma,p}(\cdot)\right)=\nonumber
\, \sum_{x_1\cdots x_t\in A^t}p(x_1\cdots
x_t)\,\,\rho_{\gamma,p}(x_1\cdots x_t).
$$
It is known that the error of Laplace predictor goes to 0 for any
Bernoulli  source $p$. More precisely, it is proven that
\begin{equation}\label{rL} \rho^t(p\|L)
< (|A| - 1)/ (t + 1) \end{equation}
 for any source $p;\,$ \cite{Ry2, RT}.

Obviously, the convergence to 0 of a predictor's error for any
source from some set $M$  is an important property. For example,
we can see from (\ref{rL}) that it is true for the Laplace
predictor and the set of  Bernoulli  sources $M_0(A)$.
Unfortunately, it is known that a predictor, which error
(\ref{r0}) goes to 0 for any stationary and ergodic source, does
not exist.
 More precisely, for any predictor
$\gamma$ there exists such a stationary and ergodic source
$\tilde{p},$ that $ \lim_{t\rightarrow\infty} \sup\: (\:\,
\rho_{\gamma,\tilde{p}}(x_1\cdots x_t)\,\: \geq const > 0 $ with
probability 1;  \cite{Ry1}. (See also \cite{Al,Mo,No}, where this
result is generalized and a history of its discovery is described.
In particular, they found out that such a result was described by
Bailey \cite{B} in his unpublished thesis). That is why it is
difficult to use (\ref{r0})
 for comparison of different predictors. On the other
hand, it is shown in \cite{Ry0,Ry1} that there exists such a
predictor $R$, that the following average $
  t^{-1}\: \sum_{i=1}^t
\rho_{R,p} (x_1\cdots x_t) $
 goes to 0 (with probability 1 ) for any stationary and ergodic source
$p,$ where  $t$ goes to infinity.  That is why we will focus our
attention on such averages. First, we define for any  predictor
$\pi$ the following probability distribution
$$ \pi(x_1 \ldots x_t) = \prod_{i=1}^t \pi(x_i | x_1 ...
x_{i-1}) . $$ For example, we obtain for the Laplace predictor $L$
that $L(0101) = \frac{1}{2} \frac{1}{3} \frac{1}{2} \frac{2}{5} =
\frac{1}{30},$ see (\ref{L}). Then, by analogy with (\ref{r0}) we
will estimate the error by K-L divergence  and define
\begin{equation}\label{R0} \rho_{\gamma, p}(x_1 ... x_t) =
t^{-1}\: ( \log ( p (x_1 ... x_t) / \gamma (x_1 ... x_t) )
\end{equation} and
\begin{equation}\label{RR} \bar{\rho}_t (\gamma, p) = t^{-1} \sum_{x_1 ... x_t \in A^t} p(x_1 ... x_t)
\log ( p (x_1 ... x_t) / \gamma (x_1 ... x_t) ).
\end{equation}
For example, from those definitions and (\ref{rL}) we obtain the
following estimation for Laplace predictor $L$ and any Bernoulli
source: $ \bar{\rho}_t (L, p) < ( (|A| - 1 ) \log t  + c)/t, $
where $c$ is a constant.

The universal predictors will play a key rule in suggested below
tests. By definition, a predictor $\gamma$ is called a universal
(in average) for a class of sources $M$, if for any $p \in M$ the
error $\bar{\rho}_t (\gamma, p)$ goes to 0, where $t$ goes to
infinity. A predictor $\gamma$ is called universal with
probability 1, if the error $\rho_{\gamma, p}(x_1 ... x_t)$ goes
to 0 not only in average, but for almost all sequences $x_1 x_2
... $. For short, we will say that the predictor (or probability
distribution) $\gamma $ is universal, if $
\lim_{t\rightarrow\infty}\rho_{\gamma, p}(x_1 ... x_t) = 0$  is
valid with probability 1  for any stationary and ergodic source
(i.e. for any $p \in M_\infty(A)$). Now there are quite many known
universal predictors. One of the first such predictors is
described in \cite{Ry0}.

\textbf{2.1 Universal coding.}

This short subparagraph is intended to give some
explanation about why and how methods of data compression can be
used for testing of independence. The point is that the prediction
problem is deeply connected with the theory of universal coding.
Moreover, practically used data compression methods (or so-called
archivers) can be directly applied for testing.

Let us give some definitions. Let, as before,  $A$ be a finite
alphabet and,  by definition, $A^* =\bigcup_{n=1}^\infty A^n $ and
$A^\infty$ is the set of all infinite words $x_1x_2 \ldots $ over
the alphabet $A$. A data compression method (or code) $\varphi$ is
defined as a set of mappings $\varphi_n $ such that $\varphi_n :
A^n \rightarrow \{ 0,1 \}^*,\, n= 1,2, \ldots\, $ and for each
pair of different words $x,y \in A^n \:$ $\varphi_n(x) \neq
\varphi_n(y) .$ Informally, it means that the code $\varphi$ can
be applied for compression of each message of any length $n, n
> 0 $ over alphabet $A$ and the message can be decoded if
its code is known. One more restriction is required in Information
Theory. Namely, it is required that each sequence
$\varphi_n(x_1)\varphi_n(x_2) ...\varphi_n(x_r), r \geq 1,$ of
encoded words from the set $A^n, n\geq 1,$ can be uniquely decoded
into $x_1x_2 ...x_r$. Such codes are called uniquely decodable.
For example, let $A=\{a,b\}$, the code $\psi_1(a) = 0, \psi_1(b) =
00, $  obviously, is not uniquely decodable. (Indeed, the word
$000$ can be decoded in both $ab$ and $ba.$) It is well known
 that if  a code $\varphi$ is uniquely decodable
then  the lengths of the codewords satisfy the following
inequality (the Kraft inequality):
$$
\Sigma_{u \in A^n}\: 2^{- |\varphi_n (u) |} \leq 1\:,$$

see, for ex., \cite{Ga}. It will be convenient to reformulate this
property as follows:

\textbf{Claim 1.} Let $\varphi$ be a uniquely decodable code over
an alphabet $A$. Then  for any integer $n$ there exists a measure
$\mu_\varphi$ on $A^n$ such that
\begin{equation}\label{kra}
- \log \mu_\varphi (u) \leq |\varphi (u)|
\end{equation} for any $u$ from $A^n.$
(Obviously, it is true for the measure $\mu_\varphi (u) = 2^{-
|\varphi (u) |} / \Sigma_{u \in A^n}\: 2^{- |\varphi (u) |}.$) It
is known in Information Theory that sequences $x_1 ... x_t,$
generated by a stationary and ergodic source $p,$ can be
"compressed" till the length $- \log
 p(x_1 ... x_t)$ bits.   There exist  so-called
universal codes, which, in a certain sense, are the best
"compressors" for all stationary and ergodic sources. The formal
definition is as follows: A code $\varphi$ is universal if for any
stationary and ergodic source $p$
$$ \lim_{t \rightarrow \infty} t^{-1} (- \log
 p(x_1 ... x_t) - |\varphi(x_1 ... x_t)|  \, = \,0 $$
 with  probability 1. So, informally speaking, the universal
 codes estimate the probability characteristics of the source $p$ and
 use them for efficient "compression".

\section{The Tests.}

In this paragraph we describe the suggested tests.
 First, we give
some definitions. Let $v$ be a  word $v = v_1 ... v_k, k \leq t,
v_i \in A .$ Denote the rate of a word $v$ occurring in the
sequence $x_1 x_2 \ldots x_k$ , $x_2x_3 \ldots x_{k+1}$, $x_3x_4
\ldots x_{k+2}$, $ \ldots $, $x_{t-k+1} \ldots x_t$ as $\nu^t(v)$.
For example, if $x_1 ... x_t = 000100$ and $v= 00, $ then $
\nu^6(00) = 3$. Now we define for any $k \geq 0$ a so-called
empirical  Shannon entropy of  order $k$ as follows:
\begin{equation}\label{He}
h^*_{ k}( x_1 \ldots x_t) = \end{equation} $$-
\frac{1}{(t-k)}\sum_{v \in A^k} \bar{\nu}^t(v)  \sum_{a \in A}
(\nu^t(va) / \bar{\nu}^t(v)) \log (\nu^t(va) / \bar{\nu}^t(v))\, ,
$$ where $k < t$ and $ \bar{\nu}^t(v  )= \sum_{a \in A} \nu^t(v a
). $ In particular, if $k=0$, we obtain
$$ h^*_{ 0}( x_1 \ldots x_t) = -  \frac{1}{t} \sum_{a \in A} \nu^t(a)  \log
(\nu^t(a) / t )\, , $$

The suggested test is as follows.

\emph{Let $\sigma$ be any probability distribution over $A^t.$ By
definition, the hypothesis $H_0$  is accepted if
\begin{equation}\label{cr}
 (t-m) h^*_{m}(x_1 ... x_t) -  \log (1/\sigma(x_1 ... x_t) ) \leq
\log (1 / \alpha) \,,
\end{equation} where $\alpha \in (0,1) .$  Otherwise, $H_0$ is rejected.}
 We denote this
test by $\Upsilon_{\alpha,\,\sigma, m}^{t}.$

\textbf{Theorem.} i) For any probability distribution (or
predictor) $\sigma$ the  Type I error of the test
$\Upsilon_{\alpha,\,\sigma, m}^{t}$ is less than or equal to
$\alpha, \alpha \in (0,1)$.

ii) If $\sigma$ is a universal predictor (measure) (i.e., by
definition, for any $p \in M_\infty (A)$
\begin{equation}\label{un} \lim_{t \rightarrow \infty} t^{-1}  (- \log
 p(x_1 ... x_t) - \log (1 /\sigma(x_1 ... x_t) ) \, = \,0 \end{equation} with probability
1), then the  Type II error goes to 0, where $t$ goes to infinity.

The proof is given in Appendix.

\textbf{Comment.} Let $\varphi$ be a uniquely decodable code (or a
data compression method). Define the test
$\hat{\Upsilon}_{\alpha,\,\varphi, m}^{t}$ as follows: The
hypothesis $H_0$  is accepted if
\begin{equation}\label{cr}
 (t-m) h^*_{m}(x_1 ... x_t) -  |\varphi(x_1 ... x_t)| \leq
\log (1 / \alpha) \,,
\end{equation} where $\alpha \in (0,1) .$  Otherwise, $H_0$ is rejected.

We immediately obtain from the theorem 1 and the claim 1 the
following statement.

\textbf{Claim 2.}  i) For any uniquely decodable code $\varphi$
the  Type I error of the test $\hat{\Upsilon}_{\alpha,\,\varphi,
m}^{t}$ is less than or equal to $\alpha, \alpha \in (0,1)$.

ii) If $\varphi$ is a universal code, then the  Type II error goes
to 0, where $t$ goes to infinity.

\section{Conclusion.}
The described above tests can be based on known universal codes
(or so-called archivers) which are used for text compression
everywhere. It is important to note that, on the one hand, the
universal codes and archivers are based on  results of Information
Theory, the theory of algorithms and some other branches of
mathematics; see, for example, probability \cite{e, Ki, K-Y, Ri,
SS}. On the other hand, the archivers have shown high efficiency
in practise as compressors of texts, DNA sequences and many other
types of real data. In fact, the archivers can find many kinds of
latent regularities, that is why they  look like a promising tool
for independence testing and its generalizations.

The natural question is a possibility of generalization of the
suggested tests for a case of an infinite source alphabet $A$
(say, $A$ is a metric space.) Apparently, such a generalization
can be done for a case of independence testing, if we will use
known methods of partitioning; \cite{DV1,DV}. But we do not know
how to generalize the suggested tests for a case where  $H_0$ is
that the source is Markovian. The point is that the partitioning
can increase the source memory. For example, even if the alphabet
$A$ contains three letters and we combine two of
 them in one subset (i.e. a new letter) the memory of the obtained
 source can increase till infinity. Hence, the generalization
  to  Markov sources with infinite alphabet can be considered as an
 open problem.

\newpage
\section{Appendix.}

\textit{Proof} of Theorem. First we show that for any Bernoulli
source   $\tau^*$ and any word $x_1 \ldots x_t \in A^t, t > 1, $
the following inequality is valid:
\begin{equation}\label{ta}
\tau^* (x_1 \ldots x_t) = \prod_{a \in A} \tau(a)^{\nu^t(a)} \leq
\prod_{a \in A} ( \nu^t(a)/t)^{\nu^t(a)} \end{equation} Indeed,
the equality is true, because $\tau^*$ is a Bernoulli measure. The
inequality follows from the well known inequality $\sum_{a \in A}
p(a) \log (p(a)/ q(a)) \geq 0, $ for K-L divergence,  which is
true for any distributions $p$ and $q$ (see, for ex., \cite{Ga}).
So, if $ p(a) = \nu^t(a)/t$ and $ q(a) = \tau^* (a),$ then
$$ \sum_{a \in A} \frac{\nu^t(a)}{t} \log \frac{(\nu^t(a)/t)}{\tau(a) }\geq
0.
$$ From the last inequality we obtain (\ref{ta}).

Let now $\tau$ belong to $M_m(A), m > 0.$ We will prove that for
any $x_1 \ldots x_t $
\begin{equation}\label{taa}
\tau (x_1 \ldots x_t) \leq \prod_{u \in A^m } \prod_{a \in A} (
\nu^t(ua)/\bar{\nu}^t(u))^{\nu^t(ua)} \:. \end{equation} Indeed,
we can present $\tau (x_1 \ldots x_t)$ as
$$\tau (x_1 \ldots x_t)= \tau_\infty (x_1 \ldots x_m) \prod_{u \in A^m } \prod_{a \in A} \tau (a/u)^{\nu^t(ua)}\:,$$
where $\tau_\infty (x_1 \ldots x_m)$ is the limit probability of
the word $x_1 \ldots x_m .$ From the last equality we can see that
$$\tau (x_1 \ldots x_t) \leq \prod_{u \in A^m } \prod_{a \in A} \tau (a/u)^{\nu^t(ua)}\:.$$
Taking into account the inequality (\ref{ta}), we obtain
$$\prod_{a \in A} \tau (a/u)^{\nu^t(ua)} \leq  \prod_{a \in A} (
\nu^t(ua)/\bar{\nu}^t(u))^{\nu^t(ua)}$$ for any word $u$. So, from
the last two  inequalities we obtain (\ref{taa}).

It will be  convenient to define an auxiliary measure on $A^t$ as
follows:
\begin{equation}\label{ro}
\pi_m (x_1 ... x_t) = \Delta \:2^{- t\, h^*_m(x_1 ... x_t)}\,,
\end{equation} where $x_1 ... x_t \in A^t $ and
$$\Delta = (\sum_{x_1 ... x_t \in A^t}\,2^{- t \,h^*_m (x_1 ...
x_t)}\,)^{- 1}\,.$$ If we take into account that $$2^{- (t-m) \,
h^*_m(x_1 ... x_t)}\, = \prod_{u \in A^m } \prod_{a \in A} (
\nu^t(ua)/\bar{\nu}^t(u))^{\nu^t(ua)} \:,$$  we can see from
(\ref{taa}) and (\ref{ro}) that, for any  measure $\tau \in
M_m(A)$ and any $x_1 \ldots x_t \in A^t,   $
\begin{equation}\label{r11} \tau (x_1 \ldots x_t) \leq \pi_m  (x_1 ... x_t) / \Delta \:.
\end{equation}
Let us denote the critical set of the test
$\Upsilon_{\alpha,\,\sigma, m}^{t} $ as $C_\alpha$ i.e., by
definition,
$$ C_\alpha = \{ x_1 \ldots x_t :\; (t - m)\: h^*_m(x_1 \ldots x_t) - \log (1/\sigma(x_1 ... x_t) )
  $$ \begin{equation}\label{C}    >
\log (1 / \alpha)   \} .\end{equation}   From (\ref{r11}) and this
definition we can see that for any  measure $\tau \in M_m(A)$
\begin{equation}\label{r1}
\tau (C_\alpha) \leq \pi_m  (C_\alpha) / \Delta \:.
\end{equation}
From the definitions (\ref{C}) and (\ref{ro}) we obtain
$$ C_\alpha = \{ x_1 \ldots x_t :\;  2^{\,(t-m)\: h^*_m(x_1 \ldots x_t)} >
(\alpha \,\:\sigma(x_1 \ldots x_t) )^{-1}  \} $$
$$ = \{ x_1 \ldots x_t :\; ( \pi_m(x_1 \ldots x_t ) / \Delta )^{-1}   >
(\alpha \,\:\sigma(x_1 \ldots x_t) )^{-1} \} \:.        $$
Finally,
\begin{equation}\label{C1} C_\alpha = \{ x_1 \ldots x_t :\; \sigma(x_1 \ldots
x_t) > \pi_m(x_1 \ldots x_t ) / (\alpha\, \Delta ) \} .
\end{equation}
The following chain of inequalities and equalities is valid:
$$ 1 \geq \sum_{x_1 \ldots x_t \in C_\alpha} \sigma( x_1 \ldots
x_t) \geq \sum_{x_1 \ldots x_t \in C_\alpha} \pi_m(x_1 \ldots x_t
) / (\alpha\, \Delta )$$ $$ = \pi_m(C_\alpha)/ (\alpha\, \Delta )
\geq \tau(C_\alpha) \Delta / (\alpha\, \Delta ) = \tau(C_\alpha)
 / \alpha .$$ (Here both equalities and the first inequality are
 obvious, the second inequality  and the third one follow from
(\ref{C1}) and (\ref{r1}), correspondingly.) So, we obtain that
$\tau(C_\alpha) \leq \alpha $ for any measure $\tau \in M_m(A).$
Taking into account that $C_\alpha$ is the critical set of the
test, we can see that the probability of the  Type I error is not
greater than $\alpha.$ The first claim of the theorem is proven.

The proof of the second  statement of the theorem  will  be based
on some results of Information Theory. The $t-$ order conditional
Shannon entropy is defined as follows:
 $$ h_t(p) =     -
\sum_{x_1 ... x_t \in A^t} p(x_1 ... x_t) $$
\begin{equation}\label{SHE} \sum_{a \in A} p(a/x_1 ... x_t) \log
p(a/x_1 ... x_t),
\end{equation} where $p \in M_\infty(A).$ It is known that for
any $p \in M_\infty(A)$ firstly, $\log |A| \geq h_0(p) \geq h_1(p)
\geq ... ,$ secondly, there exists the following limit Shannon
entropy $h_\infty (p) = \lim_{t\rightarrow\infty} h_t(p) $,
thirdly, $\lim_{t\rightarrow\infty} - t^{-1} \log p(x_1 ... x_t) =
h_\infty (p) $ with the probability 1 and, finally,
 $h_m(p)$ is strictly greater than $h_\infty(p), $ if
the memory of $p$ is larger $m$, (i.e. $p \in M_\infty(A)
\setminus M_m(A)$), see, for example, \cite{Billingsley, Ga}.

Taking into account the definition of the universal predictor (see
(\ref{un})), we obtain from the above described properties of the
entropy that
\begin{equation}\label{unh}
\lim_{t\rightarrow\infty} - t^{-1} \log \sigma(x_1 ... x_t) =
h_\infty (p) \end{equation} with probability 1. It can be seen
that $h^*_m$ (\ref{cr}) is a consistent estimate for the $m-$order
Shannon entropy  (\ref{SHE}), i.e. $\lim_{t\rightarrow\infty}
h^*_m (x_1 \ldots x_t ) = h_m(p)$ with probability 1; see
 \cite{Billingsley, Ga}. Having taken into account
that $h_m(p) > h_\infty (p)$ and (\ref{unh}) we obtain from the
last equality that $\lim_{t\rightarrow\infty} ( (t - m) \,h^*_m
(x_1 \ldots x_t )  - \log (1/ \sigma (x_1 ... x_t))) = \infty .$
This proves the second statement of the theorem.


\newpage
\begin{thebibliography}{99.}

\bibitem{Al}
 Algoet, P., 1999. Universal Schemes for Learning the Best Nonlinear
    Predictor Given the Infinite Past and Side Information.
  IEEE Trans. Inform. Theory. 45  1165-1185.

\bibitem{B}
Bailey D. H. { \it Sequential schemes for classifying and
predicting ergodic processes }, PhD Dissertation, Stanford
University, 1976.

\bibitem{Billingsley} Billingsley P., 1965. Ergodic theory and information,
 John Wiley \& Sons.

\bibitem{CS}
Csisz$\acute{a}$r I., Shields P., 2000, The consistency of the BIC
Markov order estimation. Annals of Statistics, v. 6, pp.
1601-1619.


\bibitem{DV1} Darbellay G.A., Vajda I., 1998. Entropy expressions
for multivariate continuous distributions. Research Report no
1920, UTIA, Academy of Science, Prague (library@utia.cas.cz).

\bibitem{DV} Darbellay G.A., Vajda I., 1999. Estimatin of the
mutual information with data-dependent partitions. IEEE Trans.
Inform. Theory. \textbf{48}(5), 1061-1081.


\bibitem{e} Effros, M., Visweswariah, K., Kulkarni, S.R., Verdu,
S., Universal lossless source coding with the Burrows Wheeler
transform. IEEE Trans. Inform. Theory. 45,  1315-1321.




\bibitem{FE}
Feller W., 1970. An Introduction to Probabability Theory and Its
Applications, vol.1. John Wiley \& Sons, New York.


\bibitem{Ga}
Gallager R.G., 1968.  Information Theory and Reliable
Communication.  John Wiley \& Sons, New York, 1968.

\bibitem{GKR}
Ghoudi K.,  Kulperger R.J., Remillard B., 2001.  A Nonparametric
Test of Serial Independence for Time Series and Residuals. Journal
of Multivariate Analysis, 79(2), pp. 191-218.



\bibitem{Ki}
Kieffer J., 1998.   Prediction and Information Theory, Preprint,
(available at
ftp://oz.ee.umn.edu/users/kieffer/papers/prediction.pdf/ )

\bibitem{K-Y}
Kieffer, J.C., En-Hui Yang, 2000. Grammar-based codes: a new class
of universal lossless source codes. IEEE Transactions on
Information Theory ,46 (3), 737 - 754.

\bibitem{Mo}
Morvai G.,  Yakowitz S.J., Algoet P.H., 1997.  Weakly convergent
nonparametric forecasting of stationary time
     series. IEEE Trans. Inform. Theory, 43, 483 -
     498.


\bibitem{No}
 Nobel A.B., 2003.  On optimal sequential prediction. IEEE Trans.
Inform. Theory, 49(1),  83-98.

\bibitem{Ri}
Rissanen J., 1984.  Universal coding, information, prediction, and
estimation.  IEEE Trans. Inform. Theory, 30(4) 629-636.

\bibitem{Ry0}
 Ryabko B.Ya., 1984. Twice-universal coding. Problems of
Information Transmission, 20(3) 173-177.


\bibitem{Ry1}
Ryabko B.Ya., 1988. Prediction of random sequences and universal
coding. Problems of Inform. Transmission, 24(2) 87-96.

\bibitem{Ry2}
Ryabko B.Ya., 1990. A fast adaptive coding algorithm. Problems of
Inform. Transmission, 26(4) 305--317.

\bibitem{R-M}
 Ryabko B.,  Monarev V., 2005. Using Information Theory Approach to  Randomness
 Testing, v. 133(1) 95-110.


\bibitem{RT}
Ryabko B., Topsoe F., 2002.  On Asymptotically Optimal Methods of
Prediction and Adaptive Coding for Markov Sources. Journal of
Complexity, 18(1) 224-241.

\bibitem{SS}
 Savari S. A., 2000. A probabilistic approach to some
asymptotics in noiseless communication. IEEE Transactions on
Information Theory 46(4): 1246-1262.





\end{thebibliography}
\end{document}